\documentclass{JINST}
\usepackage{units}
\usepackage{gensymb}
\usepackage{booktabs}
\usepackage{gensymb}
\usepackage[margin=0.5cm]{subcaption}
\usepackage{lineno}

\title{Development of CMOS pixel sensors for tracking and vertexing in high energy physics experiments}

\author{Serhiy Senyukov$^a$\thanks{Corresponding author.},
J\'{e}r\^{o}me Baudot$^a$, Auguste Besson$^a$, Giles Claus$^a$, Lo\"{\i}c Cousin$^a$, Wojciech Dulinski$^a$, Mathieu Goffe$^a$, Boris Hippolyte$^a$, Robert Maria$^a$, Levente Molnar$^a$, Xitzel S\'{a}nchez Castro$^a$ and Marc Winter$^a$\\
\llap{$^a$} Institut Pluridisciplinaire Hubert Curien (IPHC), Universit\'{e} de Strasbourg, CNRS-IN2P3,\\
  23 Rue du Loess 67037 Strasbourg, France\\
E-mail: \email{Serhiy.Senyukov@cern.ch}}

\abstract{
CMOS pixel sensors (CPS) represent an emerging technological approach to charged particle detectors. CMOS processes allow to integrate the sensitive volume and the readout electronics of the pixel detector in a single silicon die allowing for small pixel pitch ($\lesssim \unit[20]{\mu m}$) and low material budget ($\sim$ 0.2-0.3\% $X_0$ per layer). These characteristics make CPS an attractive option for vertexing and tracking systems of high energy physics experiments requiring high spatial resolution. Moreover, thanks to the use of commercial CMOS processes for the CPS fabrication, the construction cost can be significantly reduced in comparison to the custom technologies used so far. However, the attainable performance level of the CPS in terms of radiation hardness and readout speed is mostly determined by the fabrication parameters of the CMOS processes available on the market rather than by the CPS intrinsic potential itself.  

The constant evolution of commercial CMOS processes towards smaller feature sizes and high resistivity epitaxial layers enhances steadily the radiation hardness and allows for the implementation of more complex, accelerated readout circuits. The \emph{TowerJazz} \unit[0.18]{$\mu m$} CMOS process, being one of the most relevant examples, recently became of interest for several future detector projects. The most imminent of them is an upgrade of the Inner Tracking System (ITS) of the ALICE detector at LHC. It will be followed by the Micro-Vertex Detector (MVD) of the CBM experiment at FAIR. Other experiments like ILD consider CPS as one of the viable options for flavour tagging and tracking sub-systems.

Following the successful tests of the first prototypes in 2012 proving that the radiation hardness of the technology complies with the requirements of the new ALICE-ITS and CBM-MVD, the PICSEL group of IPHC-CNRS in Strasbourg (France) started to develop two CPS (MISTRAL and ASTRAL) dedicated to these projects. The present contribution describes the current status of this development. Results of the laboratory and beam tests of several prototype chips manufactured in Spring and tested during Summer 2013 are presented.
}


\begin{document}
\section{Introduction}
Silicon pixel detectors are widely used in high energy physics experiments. In collider experiments they usually equip several cylindrical layers surrounding the collision point. Pixel detectors allow to determine the two-dimensional positions of the particles produced in a collision in each of these layers. This information is then used to reconstruct the particles' trajectories (tracks). One of the important parameters of a particle track used in physics analyses is the impact parameter, defined as the minimum distance between the track and the collision vertex. The resolution on the impact parameter $\sigma_{IP}$ can be expressed in the following way: 
\begin{equation}
\sigma_{IP}=a \oplus \frac{b}{p \cdot \sin^{3/2}\theta}
\end{equation}
as a function of the particle momentum $p$ and the track polar angle $\theta$. The parameter $a$ is proportional to the single point resolution of the pixel detectors, $\sigma_{SP}$. The parameter $b$ is proportional to the material budget of the detector. Therefore, in order to improve the impact parameter resolution it is necessary to reduce the pixel detectors' pitch to lower the $\sigma_{SP}$, and their thickness to lower the material budget. However, the space for the improvement of these parameters is rather limited in case of the hybrid pixel detectors used so far. These detectors are composed of a sensor and a readout chip interconnected via bump bonding. Technological constraints of the bump bonding process presently limit the pixel pitch to $\sim \unit[50]{\mu m}$. Moreover, due to the complex structure of the hybrid detectors comprising two silicon chips, the material budget can be only reduced down to $\sim$ 1\% $X_0$ per layer. The resulting resolution on the impact parameter is insufficient for the experiments requiring the reconstruction of short-living particles, like charmed and bottom mesons and baryons, down to the low transverse momentum range.

\section{CMOS Pixel Sensors (CPS): present status}
An idea behind CMOS pixel sensors (CPS) is to use commercial CMOS processes used for the imaging sensors production to build the readout micro-circuitry on top of the sensitive epitaxial P-layer. The charge produced in the epitaxial layer by the traversing charged particles is collected via reverse-biased diodes and is transferred to the readout chains of the top layer. Such an integration of the sensitive volume and the readout circuitry in a single silicon die alleviates the aforementioned constraints on the pixel pitch and the detector thickness. The pixel pitch for CPS is determined mostly by the space needed to implement the in-pixel circuitry and can be lowered down to $\lesssim \unit[20]{\mu m}$. In order to reduce the material budget, CPS chips can be thinned down to $\sim \unit[50]{\mu m}$. The resulting material budget can therefore reach $\sim$ 0.2-0.3\% $X_0$ per layer. As for the radiation hardness and readout speed capabilities of CPS, they are mostly determined by several fabrication parameters of the CMOS processes employed. The
first parameter is the feature size of the CMOS process. Processes with the smaller feature size are characterized by the thinner gate oxide layer providing higher tolerance to ionizing radiation. Moreover, a smaller feature size allows to implement more sophisticated readout schemes, leading to a faster readout. The second parameter is the resistivity of the epitaxial layer. A higher resistivity allows for better depletion of the epitaxial layer. Better depletion guarantees an efficient charge collection and increases the tolerance to non-ionizing radiation.

The state-of-the-art of CPS is well represented by the \emph{Ultimate}~\cite{Ultimate} chip produced using the \emph{AMS} \unit[0.35]{$\mu m$} CMOS process. This chip was developed by the PICSEL group of the IPHC-CNRS in Strasbourg (France) for the Heavy Flavour Tracker (HFT) of the STAR experiment at RHIC. 400 chips will equip the PXL~\cite{STAR-PXL} - the two innermost layers of the HFT. The \emph{Ultimate} chip features a pixel pitch of \unit[20.7]{$\mu m$} providing single point resolution better than $\unit[4]{\mu m}$. Thanks to the feature size of \unit[0.35]{$\mu m$} and the epitaxial layer with a resistivity in excess of $\sim \unit[400]{\Omega\cdot cm}$, the chip can withstand the combined radiation load of \unit[150]{kRad} and $\unit[3\times 10^{12}]{n_{eq}/cm^2}$ at the operation temperature of \unit[35] {\celsius}. In order to reduce the power dissipation, the chip is read out in the single-row rolling shutter mode with the integration time of \unit[200]{$\mu s$}. Such an integration time corresponds well to the expected interaction rate of the STAR experiment. Three out of ten sectors of the PXL where successfully installed and commissioned in May--June 2013 \cite{STAR_M}. Full system installation is planned for January 2014, followed by the first physics run in February 2014.
\section{Future projects}
Following the promising perspectives of the STAR-PXL, several other experiments became interested in using CPS. The most imminent one is ALICE~\cite{ALJINST} (A Large Ion Collider Experiment) at LHC. The Inner Tracking System (ITS) of the ALICE detector is going to be upgraded during the second long shutdown of the LHC (LS2) in 2018/19 \cite{ALICE_ITS_CDR}. The main goal of this upgrade is to allow the reconstruction of heavy flavour hadrons at low transverse momentum ($p_T<\unit[1]{GeV/c}$). This objective can be reached by improving the impact parameter resolution of the ITS by a factor of $\sim 3$. In order to provide such an improvement, the new ITS will be built comprising seven layers of CPS with a single point resolution of $\sim \unit[5-15]{\mu m}$, depending on the layer. The new detector will have to cope with the important event rates of \unit[50]{kHz} in Pb-Pb collisions and several hundreds of kHz in pp collisions. Therefore the integration time of the detector should not exceed $ \sim \unit[30]{\mu s}$. Moreover, such event rates will lead to a relatively high radiation load on the detector. According to recent simulations, the overall dose expected for the full physics program after LS2 for the innermost layer can reach up to \unit[700]{kRad} and \unit[$10^{13}$]{$n_{eq}/cm^2$} including a safety factor of ten.

Another future experiment based on CPS is the Compressed Baryonic Matter (CBM) experiment at FAIR (GSI). CPS will be used to equip the Micro Vertex Detector (MVD)~\cite{CBM-MVD}. CPS are also considered as one of the options for the inner layers of the International Large Detector (ILD)~\cite{ILD-TDR} at ILC. 

Main requirements of these future projects are summarized in Table~\ref{tab:future_projects} in comparison to those of the STAR-PXL. It is worth noting that the requirements for the radiation hardness and the readout speed of the new projects are more demanding. They couldn't all be satisfied with the \emph{AMS} \unit[0.35]{$\mu m$} CMOS process used for the fabrication of \emph{Ultimate}. Therefore, another CMOS process had to be found, which would comply with these new requirements. In 2011, the \emph{TowerJazz}\footnote{\href{http://www.jazzsemi.com/}{http://www.jazzsemi.com/}} \unit[0.18]{$\mu m$} CMOS process was proposed as a possible solution. Thanks to the reduced feature size and high resistivity epitaxial layer ($\rho > 1 k\Omega\cdot cm$) it was considered as a good candidate. R\&D activities were initiated by several groups in order to investigate this new process.
\begin{table*}
\caption{Main requirements of several future projects compared to those of the STAR-PXL.}
\label{tab:future_projects}
\centering
\begin{tabular}{cccccc}
\toprule
 & $\sigma_{sp}$ [$\mu m$] & Integration time [$\mu s$] & TID [MRad] & NIEL [$n_{eq}/cm^2$] & Temp [\celsius]\\
\midrule
\textit{STAR-PXL} & $\mathit{\sim 4}$ & \textit{200} & \textit{0.15} & $\mathit{3\times 10^{12}}$ & \textit{35} \\
\midrule
ALICE-ITS & $\sim 5$ & $\lesssim 30$ & 0.7 & $10^{13}$ & 30 \\
CBM-MVD & $\sim 5$ & 10--30 & 10 & $10^{14}$ & $<0$ \\
ILD-VXD & $\sim 3$ & $\leq 10$ & $\sim 0.1$ & $\sim 10^{11} $ & 30 \\
\bottomrule
\end{tabular}
\end{table*}

\section{Current developments by the PICSEL group}
The PICSEL group started the CPS developments for the ALICE-ITS upgrade in 2011. The MIMOSA-32/32Ter chips were designed first, aiming to assess the charge collection performance and the radiation hardness of the \emph{TowerJazz} \unit[0.18]{$\mu m$} CMOS process. Multiple laboratory and beam tests of these chips were performed in 2012. They allowed to validate the charge collection efficiency of the process. The radiation hardness of the process was confirmed up to the combined load of \unit[1]{MRad} and $\unit[10^{13}]{n_{eq}/cm^2}$ at \unit[30] {\celsius} \cite{Senyukov_RESMDD12}.

These encouraging results allowed to pursue further developments towards the full scale CPS. Presently two architectures, called MISTRAL and ASTRAL, are being developed in parallel. Their main characteristics are summarized in Table~\ref{tab:M/A_characteristics}. The principal difference between MISTRAL and ASTRAL is that the former features end-of-column discriminators, while the latter incorporates a discriminator in each pixel. The in-pixel discriminators will allow to decrease both the integration time and the power density of the chip.

\begin{table}
\caption{Main characteristics of the MISTRAL and ASTRAL architectures.}
\label{tab:M/A_characteristics}
\centering
\begin{tabular}{ccc}
\toprule
Parameter & MISTRAL & ASTRAL\\
\midrule
Pixel size ($r\phi \times z$) [$\mu m$] & $33\times 22$ & $31\times 24$\\
CDS & \multicolumn{2}{c}{in-pixel} \\
Discriminator & end-of-column & in-pixel \\
Readout mode & \multicolumn{2}{c}{Rolling shutter}\\
Nb. of rows read out in parallel & \multicolumn{2}{c}{2} \\
Integration time [$\mu s$]& 30 & 20 \\
Power density [$mW/cm^2$]& $<200$ & $\leq 100$\\
\bottomrule
\end{tabular}
\end{table}

The development of a full scale CPS requires several prototyping steps in order to validate different building blocks of the final design. For this purpose, an engineering run including 26 prototype chips was submitted in March 2013, in collaboration with groups from CERN (Geneva) and RAL (UK). Wafers featuring epitaxial layers with different thickness and resistivity were used for the chips fabrication. Most of the tests were performed on the chips with an epitaxial layer with a thickness of $\sim \unit[18]{\mu m}$ and a resistivity of $\gtrsim \unit[1]{k\Omega \cdot cm}$ (HR-18) and that with a thickness of $\sim \unit[20]{\mu m}$ and a resistivity of $\sim \unit[6]{k\Omega \cdot cm}$ (HR-20). 

\section{Preliminary prototype tests results}
The engineering run included 10 chips developed by the PICSEL group. Starting from June 2013 these chips were tested in various ways. Laboratory tests performed at IPHC (Strasbourg) allowed to evaluate the noise and charge collection efficiency (CCE). The CCE was measured using the \unit[5.9]{keV} X-rays emitted by a $^{55}$Fe source. In August 2013, several chips were also tested with an electron beam of $\sim\unit[4]{GeV/c}$ in DESY (Hamburg). In the following, preliminary test results of the three most important prototypes (called MIMOSA-34/22THR-A1/22THR-B) are presented.

\paragraph{MIMOSA-34}
This chip was designed to study the charge collection properties of the \emph{TowerJazz} \unit[0.18]{$\mu m$} process as a function of the dimensions and design of the pixel and of the properties of the epitaxial layer. It includes 30 independent submatrices of $64\times16$ pixels each. Analogue readout is performed in the rolling shutter mode, with an integration time of $\unit[32]{\mu s}$. The pixel pitch and the collection diode surface vary from one submatrix to another. The pixel dimensions vary from $22\times 27$ to $\unit[33\times 66]{\mu m^2}$. The surface of the collection diode varies from 1 to \unit[15]{$\mu m^2$}.  

\begin{figure}[tbp] 
\centering
\begin{subfigure}[tbp]{0.45\textwidth}
\includegraphics[width=\textwidth]{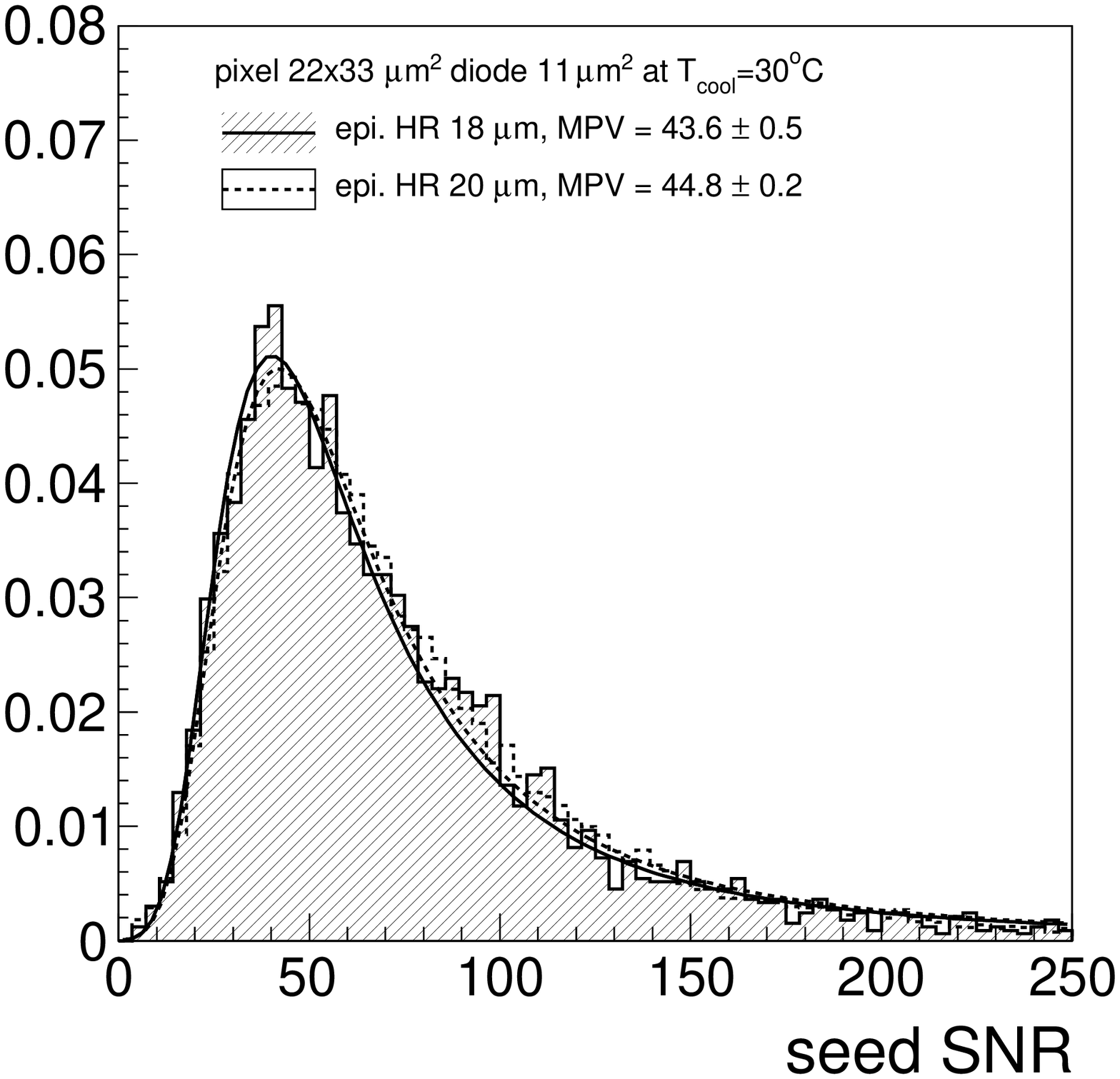}
\caption{Comparison of the pixels implemented on HR-18 and HR-20 epitaxial layers with the diode surface of \unit[11]{$\mu m^2$}.}
\label{fig:P29_vs_epi}
\end{subfigure}
\begin{subfigure}[tbp]{0.45\textwidth}
\includegraphics[width=\textwidth]{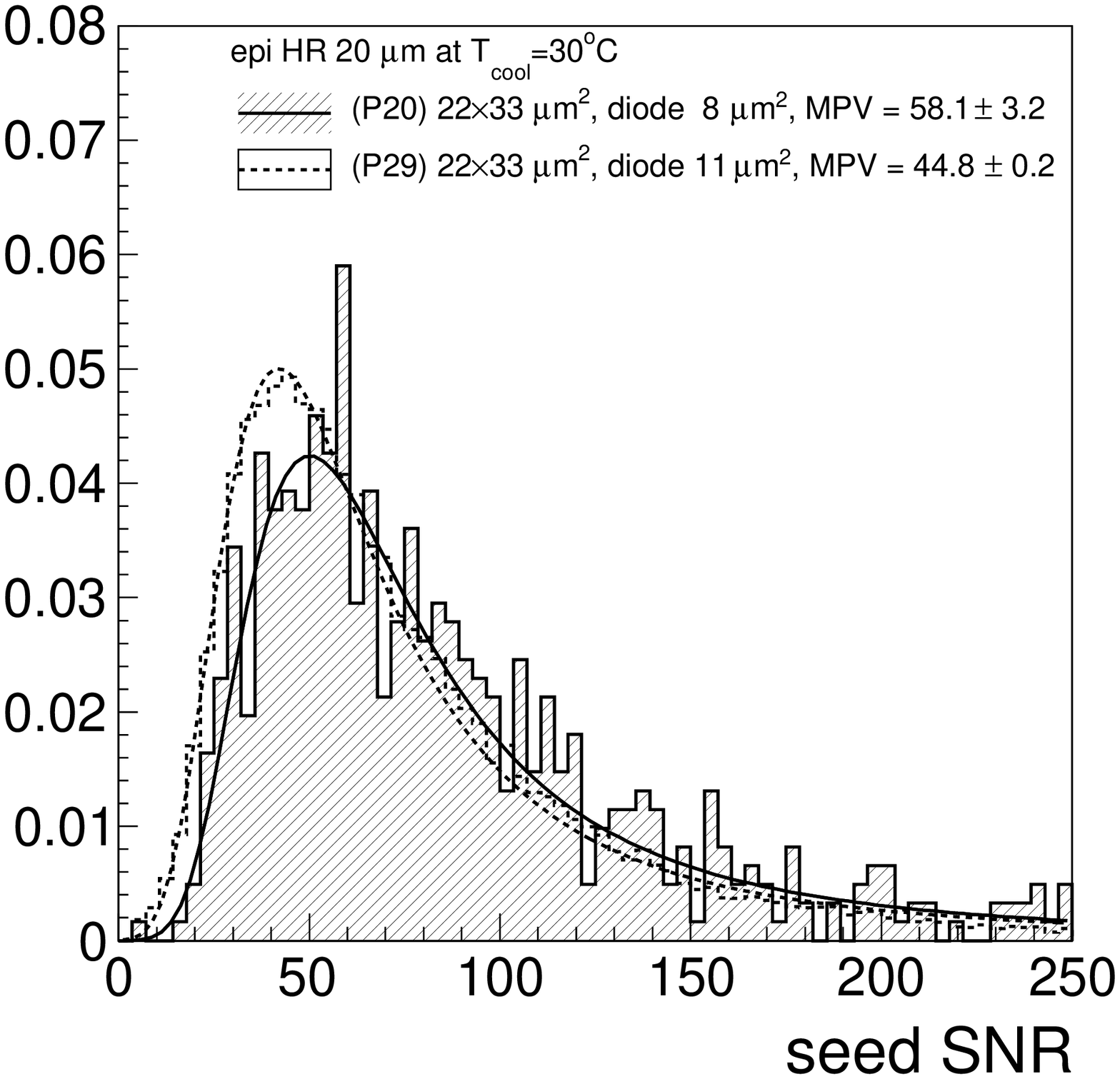}
\caption{Comparison of the pixels with the diode surface of \unit[8]{$\mu m^2$} and \unit[11]{$\mu m^2$} implemented on the HR-20 epitaxial layer.}
\label{fig:P29vsP20}
\end{subfigure}
\caption{Signal-to-noise ratio distributions of the cluster seed for different MIMOSA-34 pixels implemented on different epitaxial layers.}
\end{figure}

The signal-to-noise ratio (SNR) of the cluster seed has been measured for different submatrices and epitaxial layers during the beam test. The clusters were associated to the electron tracks reconstructed in the 6-plane pixel telescope. On Figure~\ref{fig:P29_vs_epi}, the SNR distributions of the cluster seed are compared for the pixels with the pitch of $\unit[22\times 33]{\mu m^2}$ and the diode surface of $\unit[11]{\mu m^2}$ (called P29) implemented on HR-18 and HR-20 epitaxial layers. One observes that both epitaxial layers provide similarly high SNR.

Figure \ref{fig:P29vsP20} shows the SNR distributions of the cluster seed for the pixels with the same pitch of $\unit[22\times 33]{\mu m^2}$ but different diode surfaces of 8 and \unit[11]{$\mu m^2$} (called P20 and P29), both implemented on the HR-20 epitaxial layer. One observes that the P20 pixels provide higher SNR. It can be explained by the fact that the smaller diodes of the P20 pixels having lower capacity lead to a lower pixel noise. Indeed, such an effect has been observed during the laboratory noise measurements. At the same time, it was observed that the smaller diodes of the P20 pixels provide slightly lower CCE. However, the net result of these two opposite effects is an increase of the SNR for the pixels with the smallest diode. 

\begin{figure}[tbp]
\centering
\includegraphics[width=0.5\textwidth]{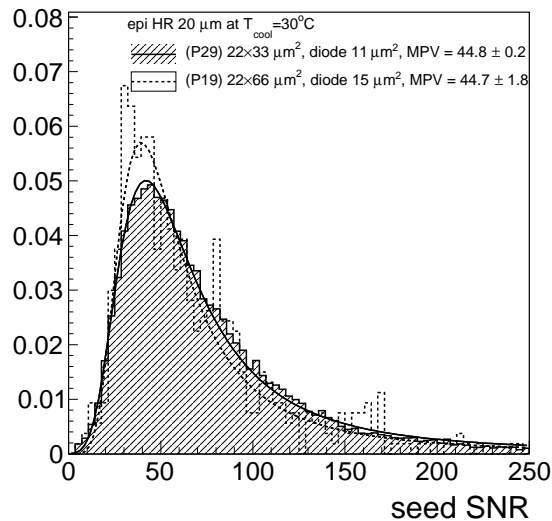}
\caption{Signal-to-noise distributions of the cluster seed for MIMOSA-34 pixels of different dimensions and diode surface implemented on the HR-20 epitaxial layer.}
\label{fig:P19vsP29}
\end{figure}

Figure \ref{fig:P19vsP29} shows the SNR distributions of the cluster seed for the pixels with dimensions of $\unit[22\times 33]{\mu m^2}$ and diode surface of \unit[11]{$\mu m^2$} and those with dimensions of $\unit[22\times 66]{\mu m^2}$ and diode surface of \unit[15]{$\mu m^2$} (called P29 and P19 respectively). One observes that despite their significantly larger pitch, P19 pixels exhibit SNR values nearly as large as those of P29. This can be explained by assuming that the larger diode of P19 compensates the charge collection efficiency loss due to the larger distance between neighbouring diodes. The results of these measurements will be used for the optimization of the pixel design of the final sensors.

Another important parameter addressed with the help of the MIMOSA-34 chip is the single point resolution. This measurement was also performed during the beam test. Due to the analogue readout of the MIMOSA-34 chip, additional data post-processing was needed, consisting in emulating of the binary discriminated output of the final chip. Table~\ref{tab:resolution} reports the preliminary values obtained with the P29 and P19 pixels ($\unit[22\times 33]{\mu m^2}$ and $\unit[22\times 66]{\mu m^2}$ pixel dimensions respectively). Results of similar measurements performed with the MIMOSA-32Ter chip are shown as well. Results obtained with the \emph{TowerJazz}-based chips are also compared to those of the \emph{AMS}-based \emph{Ultimate} chip. One observes that for similar pixel dimensions, the \emph{TowerJazz} process provides a slightly better resolution. $\unit[22\times 33]{\mu m^2}$ large pixels provide a resolution complying with the ALICE-ITS upgrade requirements. Moreover, the resolution provided by $\unit[22\times 66]{\mu m^2}$ large pixels make them a possible option for the external layers of the new ALICE-ITS, where they allow to significantly reduce the power density of the detector. 

\begin{table}
\centering
\caption{Single point resolution corresponding to the binary cluster encoding for the different pixel dimensions. The values were obtained using the eta function algorithm.}
\label{tab:resolution}
\begin{tabular}{c|c|cccc}
\toprule
Process & \emph{AMS} 0.35 $\mu m$ & \multicolumn{4}{c}{\emph{TowerJazz} 0.18 $\mu m$}\\
Chip type & Ultimate & MIMOSA-32Ter & MIMOSA-34 & MIMOSA-32Ter & MIMOSA-34 \\
Pixel size [$\mu m^2$] & $20.7 \times 20.7$ & $20 \times 20$ & $22 \times 33$ & $20 \times 40$ & $22 \times 66$ \\
\midrule
$\sigma_{sp}$ [$\mu m$] & $3.7\pm 0.1$ & $3.2\pm 0.1$ & $\sim 5$ & $5.4\pm 0.1$ & $\sim 7$ \\
\bottomrule
\end{tabular}
\end{table}

\paragraph{MIMOSA-22THR-A1} This prototype chip served several purposes. The first one was to validate the single-row rolling shutter readout via end-of-column discriminators. This architecture was used in the \emph{Ultimate} chip fabricated in the \emph{AMS} \unit[0.35]{$\mu m$} CMOS process. This time, the \emph{TowerJazz}-based implementation of this architecture had to be tested. The pixel matrix of the chip is formed by 136 columns containing 320 pixels each. 128 columns are read out in binary mode via end-of-column discriminators. 8 remaining columns provide an analogue output to test the in-pixel circuitry. The validation consisted in measuring of the particle detection efficiency and the fake hit rate as a function of the discriminators' threshold during the beam test. 

The second purpose was to find the way to reduce an excessive, non-gaussian noise observed in the MIMOSA-32/32Ter chips~\cite{Baudot_VCI2013}. A detailed noise analysis showed that this excess was due to the small fraction of pixels exhibiting RTS (Random Telegraphic Signal) fluctuations of the baseline. Several studies, e.g. \cite{RTS_2}, indicate that the magnitude of the RTS induced noise can be reduced by increasing the size of the gate of the pre-amplifier's input transistor. In order to verify this solution, the pixel array of the MIMOSA-22THR-A1 chip was composed of three distinct sub-arrays (called S1, S2 and S4) of pixels featuring a different size of the pre-amplifier's input transistor gate. In submatrix S4 the transistor gate has a length $L$ of $\unit[0.18]{\mu m}$ and a width $W$ of $\unit[1]{\mu m}$. These dimensions correspond to the pixels used in MIMOSA-32/32Ter chips \cite{Baudot_VCI2013}. Submatrix S2 contains pixels with $L=\unit[0.36]{\mu m}$ and $W=\unit[1]{\mu m}$, i.e. the length of the gate has been doubled. Submatrix S1 comprises pixels with $L=\unit[0.36]{\mu m}$ and $W=\unit[2]{\mu m}$. In this case, both the length and the width of the gate have been doubled.

\begin{figure}[tbp] 
\centering
\includegraphics[width=1.\textwidth]{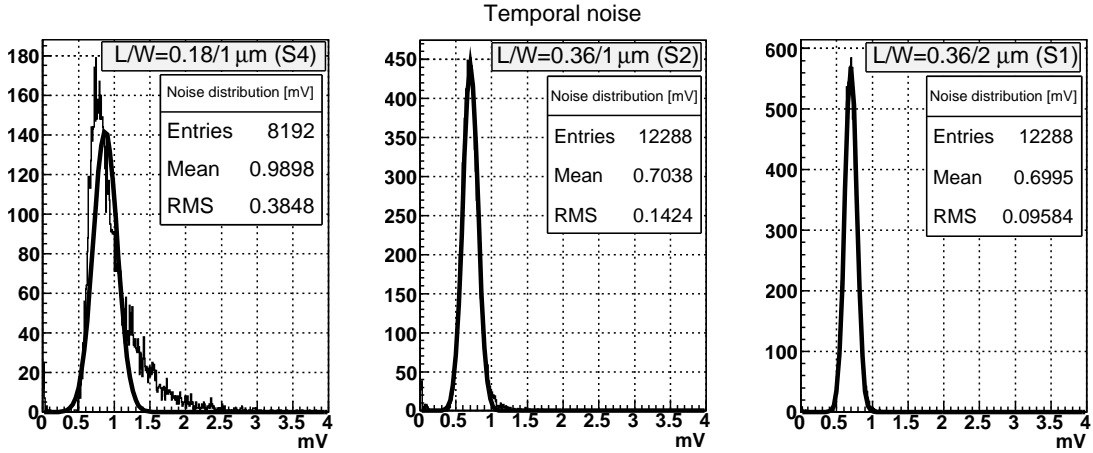}
\caption{Distributions of the temporal noise for three submatrices of the MIMOSA-22THR-A1 chip featuring different sizes of the input transistor gate (indicated at the top of each distribution).}
\label{fig:TN_S1_vs_S2_vs_S4}
\end{figure}

The temporal noise (TN) and fixed pattern noise (FPN) were evaluated via discriminator threshold scan performed in the laboratory. Figure \ref{fig:TN_S1_vs_S2_vs_S4} shows the distributions of TN for pixels composing the submatrices S4, S2 and S1. One can see that the non-Gaussian noise tail is present in the distribution corresponding to submatrix S4 (left). These pixels have the smallest size of the input transistor gate, and are therefore the most susceptible to RTS noise. The noise tail is significantly reduced in the distribution corresponding to the submatrix S2 (middle). Indeed, the  doubled length of the gate allowed to reduce the RTS noise. Finally, the noise distribution became nearly purely Gaussian for the submatrix S1. Thus, one can conclude that the RTS noise was almost completely suppressed by the simultaneous increase of the length and width of the gate of the pre-amplifier's input transistor. 

\begin{figure}[tbp] 
\centering
\includegraphics[width=.9\textwidth]{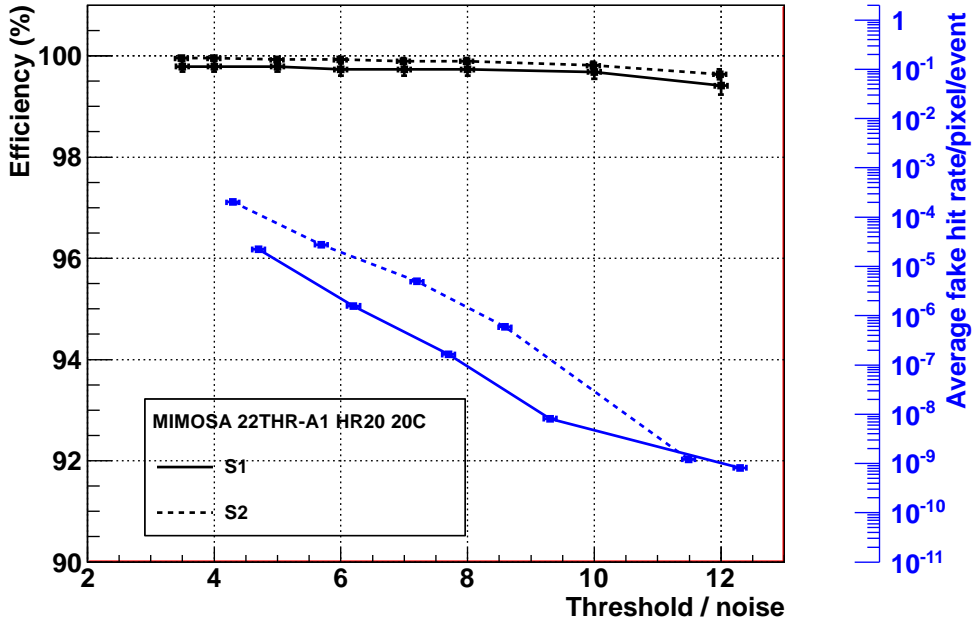}
\caption{Particle detection efficiency and fake hit rate as a function of the discriminator threshold for the submatrices S1 and S2 of the MIMOSA-22THR-A1 chip.}
\label{fig:HR20-50um-20C-S1vsS2_analog}
\end{figure}

Tests with the electron beam in DESY allowed to measure the particle detection efficiency and fake hit rate as a function of the discriminator threshold. Results of these measurements for the submatices S1 and S2 are presented in Figure \ref{fig:HR20-50um-20C-S1vsS2_analog}. One observes that the particle detection efficiency remains higher than 99.5 \% over a wide range of the discriminator thresholds. At the same time, the fake hit rate remains at an acceptable level of $\sim 10^{-5}$ or below. These results allowed to validate the single-row rolling shutter readout. Moreover, it was confirmed that by increasing the size of the gate of the input transistor one can effectively mitigate the RTS-induced noise and reduce the fake hit rate down to acceptable levels without degrading the SNR performance at a level affecting the particle detection efficiency.

\begin{figure}[tbp] 
\centering
\includegraphics[width=1.\textwidth]{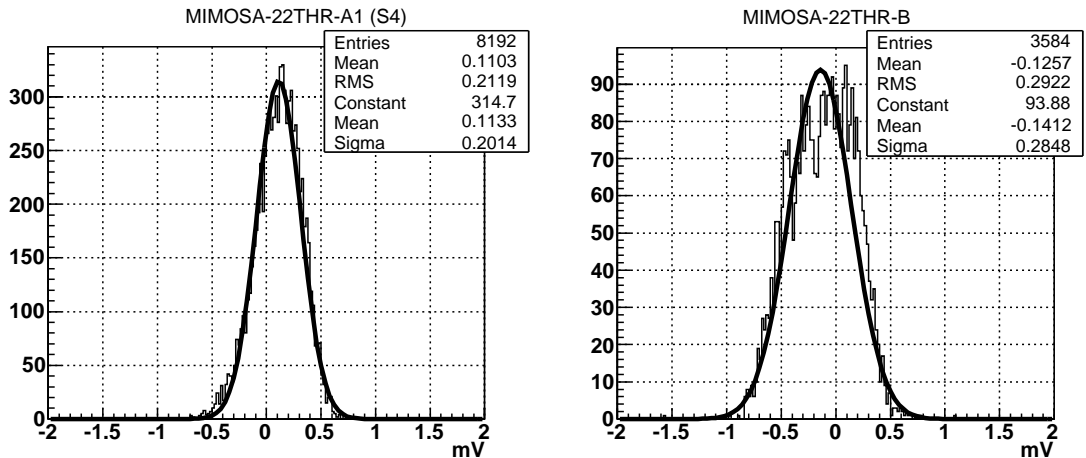}
\caption{Distributions of the fixed pattern noise for the MIMOSA-22THR-A1 (S4) on the left and MIMOSA-22THR-B on the right.}
\label{fig:FPN_A_vs_B}
\end{figure}

\paragraph{MIMOSA-22THR-B} This chip has been used to validate the double-row rolling shutter readout via end-of-column discriminators. Together with the tests of the MIMOSA-22THR-A1 it would allow validating the upstream part of the MISTRAL architecture. The pixel matrix of MIMOSA-22THR-B is composed of 64 columns made of 64 pixels each. 56 columns are read out in binary mode via end-of-column discriminators. The 8 remaining columns provide an analogue output. A threshold scan was performed in the laboratory in order to assess the TN and FPN. Figure \ref{fig:FPN_A_vs_B} compares the distributions of the FPN of MIMOSA-22THR-B (right) with that of the submatrix S4 of MIMOSA-22THR-A1 (left). One observes that the MIMOSA-22THR-B chip has higher dispersion of the noise among different pixels. This effect is still of minor consequence on the overall noise performance of the sensor. It was suspected to originate from the crosstalk between multiple lines needed for the double-row readout. This crosstalk issue has been understood and will be addressed in the next prototype chips. Apart from this issue, one can consider that the upstream part of the MISTRAL architecture has been validated.

\section{Conclusions and outlook}
Thanks to the CPS it became possible to envisage building highly granular and very light tracking and vertexing detectors for various physics experiments, the STAR-PXL being the pioneering device of this technology.

CPS are now also foreseen for the complete ITS upgrade since the \emph{TowerJazz} \unit[0.18]{$\mu m$} CMOS process has been proven to be radiation tolerant enough and a fast enough read-out architecture seems realisable. As for the CBM-MVD, the choice of CPS, which was already decided for the first stages of the experiment, seems now to be extendible to its SIS-300 physics program, which is the most demanding in terms of radiation tolerance and read-out speed. The PICSEL group of IPHC is pursuing an extensive R\&D towards the final full scale CPS for these projects. It includes two architectures to be integrated in the MISTRAL and ASTRAL chips. First tests of some prototype chips fabricated in Spring 2013 gave several promising results towards this goal. First, the upstream part of the MISTRAL architecture consisting of the double-row rolling shutter readout via end-of-column discriminators has been validated. It provides a particle detection efficiency above 99.5\%. Second, an effective noise mitigation measure has been found. It allows reducing the fake hit rate below $10^{-5}$. Finally, the single point resolution provided by the pixels with a size of $\unit[22\times 33]{\mu m^2}$ has been estimated to be $\sim \unit[5]{\mu m}$. This value satisfies the requirements of the ALICE-ITS upgrade project. These results will allow the production of the first full scale (surface $\sim \unit[1]{cm^2}$) prototype of the MISTRAL architecture in 2014. What concerns the ASTRAL architecture, it still requires additional prototyping motivated by the in-pixel noise suppression.  

\bibliographystyle{JHEP}
\bibliography{Senyukov-IPRD13-proceedings}

\end{document}